**Transport and field emission properties of buckypapers obtained from aligned carbon nanotubes**

F. Giubileo[1*], L. Iemmo[2], G. Luongo[2], N. Martucciello[1], M. Raimondo[3], L. Guadagno[3], M. Passacantando[4], K. Lafdi[5] and A. Di Bartolomeo[2,1]

[1]*CNR-SPIN Salerno, via Giovanni Paolo II, 84084 Fisciano, Italy*

[2]*Physics Department "E. R. Caianiello", University of Salerno, via Giovanni Paolo II, 84084, Fisciano, Italy*

[3]*Department of Industrial Engineering, University of Salerno, via Giovanni Paolo II, 84084, Fisciano, Italy*

[4]*Department of Physical and Chemical Science, University of L'Aquila, via Vetoio, 67100, Coppito, L'Aquila, Italy*

[5]*University of Dayton, 300 College Park Dayton Ohio 45440 USA*

The e-mail addresses of all authors:

E-mail: filippo.giubileo@spin.cnr.it -Tel:+39.089.969329 (Filippo Giubileo) *Corresponding Author

E-mail: liemmo@unisa.it (Laura Iemmo)

E-mail: g.luongo19@studenti.unisa.it (Giuseppe Luongo)

E-mail: nadia.martucciello@spin.cnr.it (Nadia Martucciello)

E-mail: mraimondo@unisa.it (MariaLuigia Raimondo)

E-mail: Lguadagn@unisa.it (Liberata Guadagno)

E-mail: maurizio.passacantando@aquila.infn.it (Maurizio Passacantando)

E-mail: klafdi1@udayton.edu (Khalid Lafdi)

E-mail: adibartolomeo@unisa.it (Antonio Di Bartolomeo)

**Abstract**

We produce 120 μm thick buckypapers from aligned carbon nanotubes. Transport characteristics evidence ohmic behavior in a wide temperature range, non linearity appearing in the current-voltage curves only close to 4.2 K. The temperature dependence of the conductance shows that transport is mostly due to thermal fluctuation induced tunneling, although to explain the whole temperature range from 4.2 K to 430 K a further linear contribution is necessary. The field emission properties are measured by means of a nanocontrolled metallic tip acting as collector electrode to access local information about buckypaper properties from areas as small as 1 μm$^2$. Emitted current up to $10^{-5}$A and turn-on field of about 140V/μm are recorded. Long operation, stability and robustness of emitters have been probed by field emission intensity monitoring for more than 12 hours at pressure of $10^{-6}$ mbar. Finally, no tuning of the emitted current was observed for in plane applied currents in the buckypaper.

**Keywords:** buckypaper; carbon nanotubes; field emission; transport properties.

**Introduction**

Since their discovery, carbon nanotubes (CNTs) [1] have been considered exceptional elements to realize field emission devices, due to their very high aspect ratio, excellent electrical conductivity and important mechanical strength. Nowadays, CNT based field emitters are used in vacuum electronics to produce electron sources [2], flat panels [3], X-ray sources [4,5], and microwave amplifiers [6], exploiting a low-threshold electric field and large emission current density. To increase the extracted current, CNT arrays are usually implemented [7,8] as free-standing well-aligned CNTs [9-11] or paper-like sheet of randomly oriented CNTs, named buckypaper [12,13].

Several reports focus on oriented samples grown on particular substrates to improve the control on dimension and spacing of CNT emitters and to get stable emitted current [14-16]. However, growth of oriented nanostructures needs carefully controlled fabrication process. The development of buckypaper field emitters is motivated by the extreme simplicity of fabrication process especially for large scale as well as ease to use. Buckypapers have a laminar structure with networks of CNTs held together by van der Waals forces. The CNTs are randomly oriented in the plane unless special techniques, as the application of electric and magnetic fields, are adopted [17] to obtain preferential alignment in one direction. The material results a good candidate for large scale emitters for the possibility to emit electrons from the whole length of the tubes [18]. Recently, buckypapers have been applied to develop supercapacitors [19,20], chemical sensors [21], flexible fibers [22] and actuators [23]. Enhancement of field emission in buckypaper has been also reported due to acid functionalization of nanotubes [24] and surface plasma treatment [19].

In this paper, we study the field emission properties of 120 µm thick buckypaper obtained by pressing aligned CNTs whose original length was up to 200 µm. The temperature dependence of the buckypaper conductance was measured in the wide temperature range 4.2K – 430K evidencing the presence of thermal fluctuation induced tunneling contribution as well as a linear contribution to the total conductance. The field emission characteristics are locally measured by using a piezoelectric driven metallic probe tip, with curvature radius of about 30 nm, in order to collect electrons emitted from areas as small as $1 \mu m^2$ of the buckypaper. We analyze the turn-on field, the emission current intensity and its time stability as well as the possibility to modify the FE current by applying an additional planar current in the sample.

**Materials and methods**

*Sample preparation*

The chemical vapor deposition (CVD) method was used to grow CNTs on a quartz wafer in a two-step process consisting of a catalyst preparation followed by the actual synthesis of the CNTs. The quartz wafer cut into appropriate size was first heated in a furnace at 500 °C for about 10 minutes. It was then cooled down to the room temperature and dipped in catalyst. The catalyst was basically a solution of ethanol and Fe-Mo in mole ratio of 10:1. The quartz substrate thus dipped in catalyst was then placed in a quartz tube in a tube furnace. The deposited Fe catalyst was then reduced at 800 °C by passing hydrogen and argon gas. Finally, the substrate was subjected to source of carbon by passing ethylene gas which caused the decomposition of carbon and resulted in the synthesis of CNTs. We managed to grow aligned CNTs with length ranging from several micrometers to about 200 µm with a narrow diameter distribution around 10 nm. Scanning electron microscope (SEM) micrographs are reported in Figure 1. Raman spectroscopy was performed by means of Renishaw Lab spectrometer, equipped with a laser excitation of 785 nm (1.58 eV excitation, 30s collection time). Measured spectra show that the obtained aligned CNTs are a mixture of single and multiwalled carbon nanotubes (Figure 1b). Subsequent steps were carried out to press the sample between two parallel plates to make the film denser and to separate it from the substrate in the form of buckypaper of about 120µm thick.

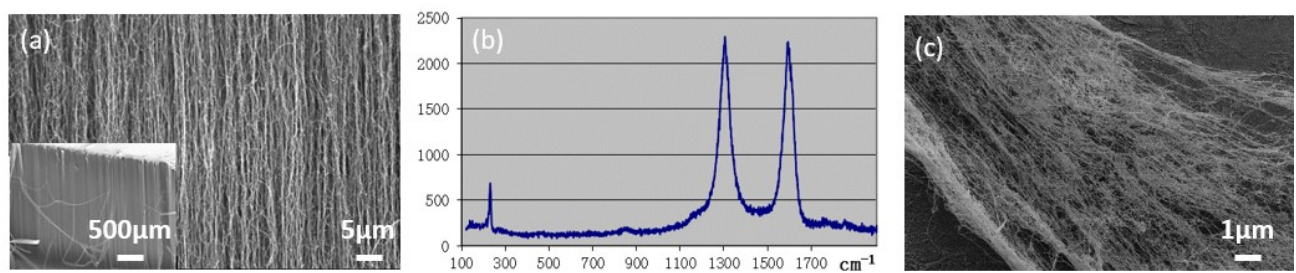

**Fig. 1 (a)** SEM images of aligned carbon nanotubes on quartz substrate. Inset: image with reduced magnification to evidence the alignment over large area. **(b)** Raman spectrum of vertically aligned CNTs with peak at 200 cm$^{-1}$ corresponding to RBM of single walled CNTs. **(c)** SEM image of the surface of as pressed 120 µm thick CNT buckypaper.

Even though our process is focused on making 100% multi-walled carbon nanotubes (MWCNTs) it produced a mixture of single-walled carbon nanotubes (SWCNTs) as well. This is shown by the 200 cm$^{-1}$ peak in the Raman spectrum reported in figure 1b. This peak corresponds to the Radial Breathing Mode (RBM), which is usually located between 75 and 300 cm$^{-1}$ [25]. The D mode (located between 1330-1360 cm$^{-1}$) and the G mode, which corresponds to the stretching mode in the graphite plane, are typical of good quality MWCNTs [26].

**Results and discussion**

*Electrical characterization*

A standard four-probe method was applied to perform the electrical characterization of the produced samples by means of Janis ST-500 Cryogenic probe station working in the temperature range from 4.2 K to 450 K and in vacuum (pressure range from $10^{-6}$ mbar to ambient atmosphere). A Semiconductor Parameter Analyzer (Keithley 4200-SCS) with four source-measurement units was connected to the probe station via triaxial cabling to perform current biased measurements, while controlling/monitoring the sample temperature.

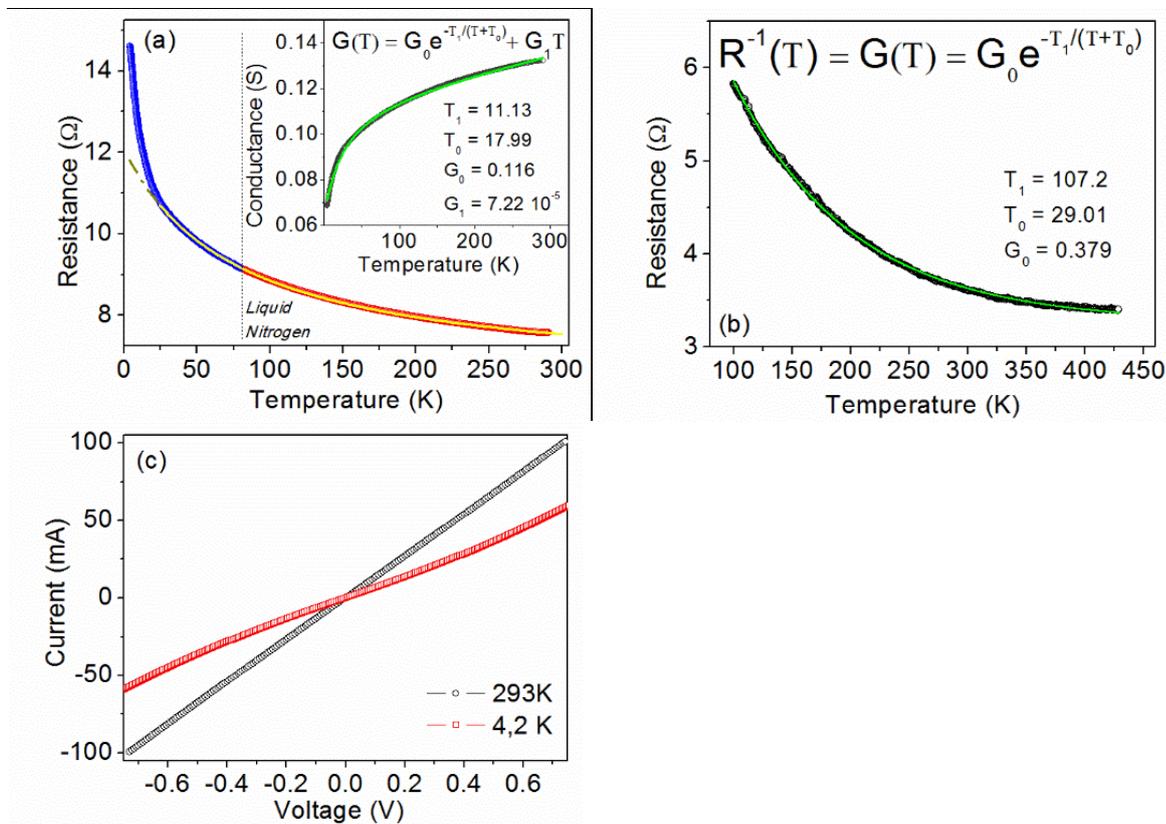

**Fig. 2 (a)** Resistance vs Temperature for a MWCNT buckypaper measured in the temperature range from 4.2 K to 300 K. Solid line (black) is the best fit by fluctuation induced tunneling model for data in the range 20 K to 300 K. Inset: the whole set of data is plotted as G=1/R and is fitted by adding to the model a linear term $G_1*T$; Solid (green) line represents the best fit. **(b)** Resistance vs Temperature for another MWCNT buckypaper measured in the temperature range from 100 K to 430 K. Solid line (green) is the best fit by fluctuation induced tunneling model. **(c)** Current-Voltage characteristics measured at room temperature and at 4.2 K.

We measured the temperature dependence of the resistance R(T) for several samples (randomly cut from the same source sample) and we observed a negative temperature coefficient of the resistance (dR/dT<0) in the whole temperature range (4.2 K to 300 K in figure 2a and 100 K to 430 K in figure 2b). Measurements were performed by forcing a low constant current (0.5 mA) to prevent sample self-heating. For MWCNT bundles [27,28] or film [29] the temperature dependence of resistivity usually remains non metallic, dR/dT<0, for the whole temperature range. A crossover temperature from metallic to non metallic has been reported in MWCNT buckypaper [30]. Differently, for SWCNTs a crossover temperature has been often observed with values varying from 35 K for a single well-ordered rope to 250 K for a rope with tangled regions [31]. It appears difficult to establish a consistent scenario for the MWCNTs also considering that below 20 K, both sharp increases in resistivity [27,32] and plateaus [28,33] have been reported for bundles and individual tubes.

From a theoretical viewpoint, for homogeneous disordered systems, the non metallic temperature dependence of the conductivity is explained in terms of variable range hopping (VRH) conduction [34], with $\sigma(T) = \sigma_0 \cdot exp\left[-\left(\frac{T_0}{T}\right)^{\frac{1}{1+d}}\right]$ where d is the dimension of the system. Alternatively, according to the thermal fluctuation induced tunneling (FIT) model [35], the conductivity is $\sigma(T) = \sigma_0 \cdot exp\left[-\frac{T_1}{T+T_0}\right]$ where $\sigma_0$ is the conductance at room temperature, $T_1$ denotes the temperature below which the conduction is dominated by the charge carrier tunneling through the barrier and $T_0$ the temperature above which the thermally activated conduction over the barrier begins to occur. FIT model has been developed for disordered heterogeneous systems such as conductor/insulator composites, granular metals, and disordered semiconductors, characterized by large conductive filaments interconnected via small insulating gaps. Due to small sizes of tunnel junction, thermal voltage fluctuation influences the electron tunneling probability through the barrier.

Electronic transport characterization measurements have sometimes shown that both VHR and FIT can give both reasonable fits; in real situations, it can be difficult to determine the dominant mechanism responsible for the observed electrical conduction.

By fitting the experimental data reported in figure 2a, we found out that the best fit over the largest temperature range ($T \geq 20\ K$) is obtained by the FIT model. For our system, the tunneling barriers originate from the inter-tubular contacts and the buckypaper can be considered a heterogeneously disordered system. For this reason, the FIT was used for describing the temperature dependence of conductivity of SWCNTs fibers [36] and networks [37] as well as for MWCNTs [38].

However, when considering our experimental data in the whole interval from 4.2 K to 300 K, neither FIT nor VRH are able to reproduce the complete behavior.

To explain temperature dependence measured down to 4.2 K on CNT mats [39,40] a further negative linear term has been sometimes introduced [41]. A linear conductance has also been observed in gas-desorbed CVD-grown MWCNTs at high temperature and it has been explained using a thermal activation picture of conduction channels. Moreover, such a linear dependence could imply a very short mean free path due to large number of defects [42]. Hence, to observe such a linear dependence, "dirty" MWCNTs with mean free path of few nanometers are necessary. Larger values up to two order of magnitude more can be obtained in "clean" MWCNTs, characterized by limited numbers of defect scatterers [43-45].

By considering a further linear contribution, we can express the total conductance as $G(T) = G_0 \cdot exp\left[-\frac{T_1}{T+T_0}\right] + G_1 T$ obtaining a perfect fit of the experimental data in the whole temperature range from 4.2 K to 300 K (inset of figure 2a).

The fitting parameters $T_1$ and $T_0$ are defined as:

$$T_1 = \frac{16\ \varepsilon_0 \hbar A V_0^{3/2}}{\pi\ e^2 k_B (2m_e)^2\ w^2}$$

$$T_0 = \frac{8\ \varepsilon_0 A V_0^2}{e^2 k_B w}$$

where $\varepsilon_0$ is the vacuum permittivity, $\hbar$ the reduced Planck constant, A the junction area, $V_0$ the height of the barrier, $w$ the width of the tunnel junction, $e$ the electronic charge, $m_e$ the electron mass, $k_B$ the Boltzmann constant. The linear correction to the total conductance is not used to fit the experimental data measured in the range (100 K – 430 K) for another sample (figure 2b). By assuming a junction area A≈d$^2$ with d≈10 nm the average MWCNT diameter, from the fitting parameters, we can estimate a junction width $w \approx 3.0$ nm and $V_0 \approx 24$ meV. $V_0$ is the tunnel barrier that electrons have to overcome to move from one nanotube to another one and the tunnel probability depends on it as well as on the local density of states of each side of the junction. The effective barrier height depends on the bias current and on the temperature [46,47]. Small $V_0$ values have been reported when, at low temperatures, current bias above 100µA are applied [48].

We also measured the current-voltage characteristics at different temperatures. For both samples, we always found a linear ohmic dependence, except for low temperatures. We report in figure 2c the curve measured at low temperature (4.2 K) for the first sample, and it is compared to the room temperature characteristic. Such non-ohmic behavior of the I-V characteristic has been observed at low temperature in SWCNT network [37,49,50] and explained by developing an electrical model that considers series-parallel connections of junctions existing in the CNT bundles [50].

*Field emission properties*

To study field emission properties we connected the Keithley 4200-SCS to a nanoprobe system manufactured by Kleindeik Company (nanomanipulators MM3A), with two piezoelectric driven arms, installed inside a Zeiss LEO 1430 SEM, allowing electrical measurements in-situ (pressure ~10$^{-6}$ mbar) by means of nanometric metallic probes (tungsten tips with 30 nm curvature radius).

In figure 3a we show the current-voltage characteristic of the buckypaper, measured by landing the two nanoprobes on macroscopic silver paint pads to favor very stable and low resistance contacts. This curve is compared to the case in which one contact is realized by pressing the metallic tip directly on the buckypaper. The experimental data confirm that silver paint is useful to reduce the contact resistance. This is not relevant for four-contacts measurements (as for temperature dependence measurements of the resistance previously reported) but it is important for the two terminal measurements, discussed in this section.

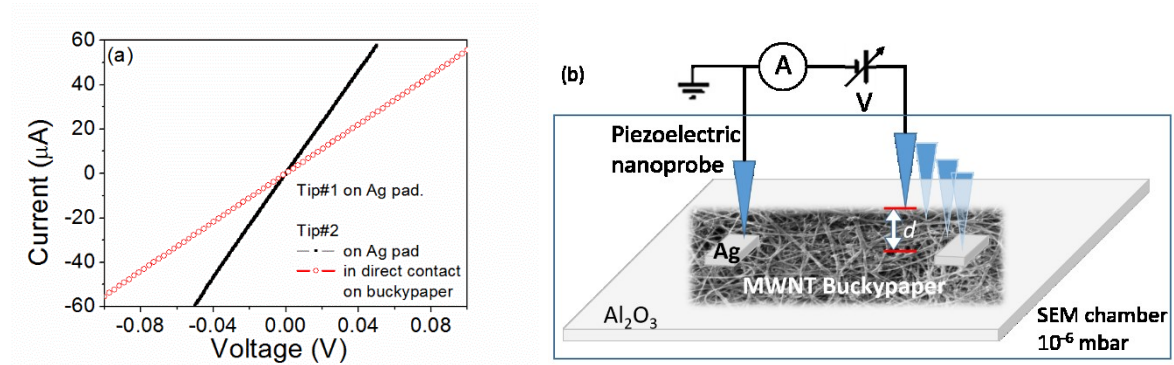

**Fig. 3 (a)** Two-probe measurement of the Current–Voltage characteristics of the buckypaper inside the SEM chamber by means of nanoprobes. Solid circles refer to configuration in which both metallic tips contact the sample on silver pads; empty squares refer to the case of one tip in direct contact with the buckypaper surface. **(b)** Schematic of the experimental setup.

The circuit configuration for field emission characterization is easily obtained by retracting one of the probes and adjusting its distance $d$ from the buckypaper surface (far from the Ag pad). A schematic of the circuit is shown in figure 3b. The piezoelectric control of the probe tips allows fine tuning of the cathode (buckypaper)-anode (metallic tip) distance with spatial resolution down to 5 nm. The use of a metallic tip as collector of the emitted electrons is a well established technique [9,10,12] that allows to get local information about the field emission properties, the electrons being emitted from a reduced area (of the order of $1 \mu m^2$) with respect the standard parallel plate setup (generally probing areas up to several $mm^2$).

To perform field emission characterization, current–voltage characteristics are measured by sweeping the voltage bias from 0 V to 120 V. Larger bias was not applied to prevent damages to the nanomanipulator circuitry. The emitted current was measured with an accuracy better than 0.1 pA.

All the field emission measurements were performed in a high vacuum (< 10⁻⁶ mbar) at room temperature.

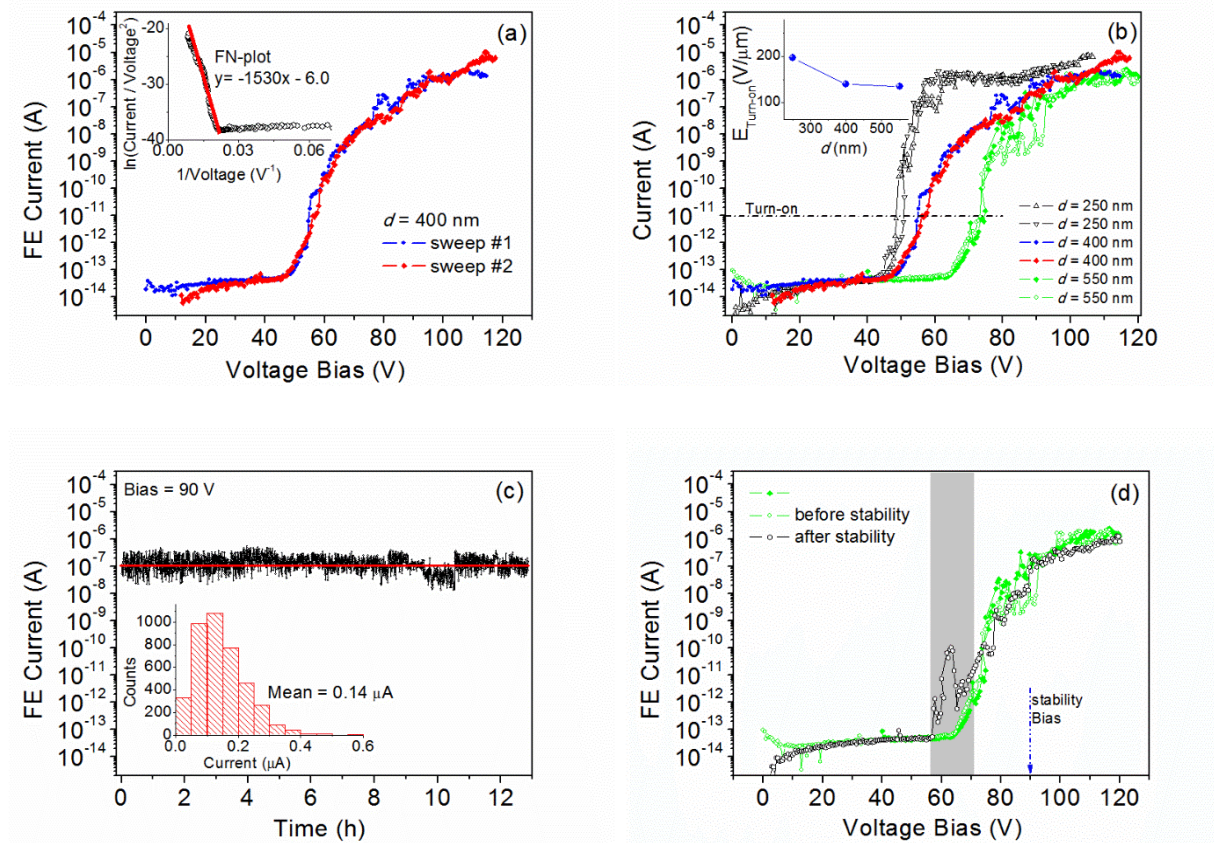

**Fig. 4 (a)** FE current vs Voltage bias characteristics measured at a tip-sample distance of 400 nm. Two successive sweeps are compared. Inset: FN-plot for the second sweep and its linear fitting. **(b)** FE curves measured for different values of the tip-sample distance d. Dashed line represents the current values considered for the evaluation of the turn-on field. Inset: Values of the turn-on field vs distance, evaluated for each curve of figure 3b. **(c)** Time stability of the FE current intensity, at fixed bias 90 V. Inset: histogram of the measured values with a rate of 1point each 12 seconds. **(d)** Comparison of the FE curves measured before and after the stability test. The gray band identifies the region in which the two curves are different.

In figure 4a we report two successively measured FE current-voltage characteristics, in the voltage bias range from 0 V to 120 V. As standard procedure, we typically perform at least two successive sweeps in order to verify the repeatability of the measurement that gives information about the robustness of the device against the electrical stress induced modifications. The experimental data of figure 4a show a current starting to flow around 50 V and exponentially growing for about eight order of magnitudes (up to several μA) from the setup floor noise of about $10^{-13}$ A. To confirm the FE nature of the measured current, we analyze the data according to the Fowler-Nordheim (FN) theory [51], which expresses the emitted current as

$$I = a \frac{E_s^2}{\varphi} S \cdot exp\left(-b \frac{\varphi^{3/2}}{E_s}\right)$$

where $\varphi$ is the work function of the CNTs, $S$ is the emitting area, $a = 1.54 \cdot 10^{-6}$ AV$^{-2}$eV and $b = 6.83 \times 10^7$ Vcm$^{-1}$eV$^{-3/2}$ are constants, and $E_s$ is the applied electric field that depends on the emitter geometry trough the field enhancement factor $\beta$ as $E_s = \beta V/d$, V being the applied bias voltage.

The factor $\beta$ represents the ratio between the local electric field on the sample surface and the applied field. According to FN-theory, for FE device we expect a linear relation in the so-called FN-plot, ln(I/V$^2$) versus 1/V, whose slope m $= k_{eff} b \varphi^{3/2} d/\beta$, can used to estimate $\beta$. In the last expression, the tip correction factor $k_{eff}$ takes into account the shape of the collector electrode [10,52]. The inset of figure 4a shows the FN-plot for the second sweep reported in the figure. The clear linear dependence demonstrate the FE nature of the recorded current. From the linear fitting we extract the field enhancement factor $\beta \approx 30$ by considering $k_{eff} \approx 1.6$ [10]. Consequently the turn-on field (E$_{Turn-ON}$ , here defined as the applied field necessary to extract a current of $10^{-11}$ A) can be evaluated as E$_{Turn-ON}$ = 140V/μm, a value significantly lower or comparable to what has been observed for other structures, such as aligned MWCNTs [10], graphene flakes [53] or nanoparticles [54].

In figure 4b we report other FE current versus bias voltage characteristics measured for different values of the tip-sample distance $d$. As expected, increasing the distance, $d$ =550 nm, the FE starts at higher applied voltages (~65V), while for reduced distance, $d$ =250 nm, the FE starts at lower applied voltages (~40V). The inset of figure 4b shows the values of the turn-on field evaluated for each distance. We notice that for $d$ =250 nm the FE characteristics follow the usual exponential growth up to a voltage of about 60 V. Above this voltage, the current is strongly limited despite the increasing bias voltage. This is probably caused by the presence of a series resistance in the circuit causing a significant voltage drop that reduces the applied field when a significant current is flowing. To study the FE current stability we used the configuration with $d$ = 550 nm and applied constant voltage bias of 90 V for measuring the FE current intensity versus time. The result is shown in figure 4c. The current has been monitored for more than 12 hours (at a rate of 1 point each 12 seconds) without significant degradation with respect the average value of 0.14 µA, demonstrating the high stability of the buckypaper with respect the long duration current emission. After the stability experiment, we repeated the complete voltage sweep (from 0 V to 120 V) to compare (see figure 4d) the FE characteristics before and after the continuous operation of the device for more than twelve hours. The curve measured after the current annealing was not affected by the long term emission almost in all the bias range. Only near the turn-on bias we observe a peak indicating an anticipated switch on of the emission process probably due to a light morphological modification, such as an elongated CNT out from the buckypaper, induced by the long exposition at high bias during the stability test. However, the high current density may burn it restoring the previous conditions. This demonstrates also the robustness and stability of buckypapers against local modifications during its operation, also better than the case of aligned CNTs, grown on Si wafer. Indeed, for aligned nanotubes, a more pronounced instability is due to the protrusion of tubes of different lengths, allowing higher current but provoking several modifications due to burning of single tubes. This is the reason why with aligned CNTs it is often necessary to perform longer electrical annealing to stabilize the emission [9-10].

According to the theoretical prediction [55] for graphene based field effect transistors [56-60], in presence of a wedge tip, enhanced local electrical field can sensibly increase the probability of electron field emission from graphene towards top gate depending on the channel current. Experimentally, a possible effect of channel current on graphene field emission [55,61] has been also reported. For this reason, we tested the possibility to tune the emitted current from the buckypaper by applying an in-plane current in the sample. The schematic of the device is reported in the inset of figure 5.

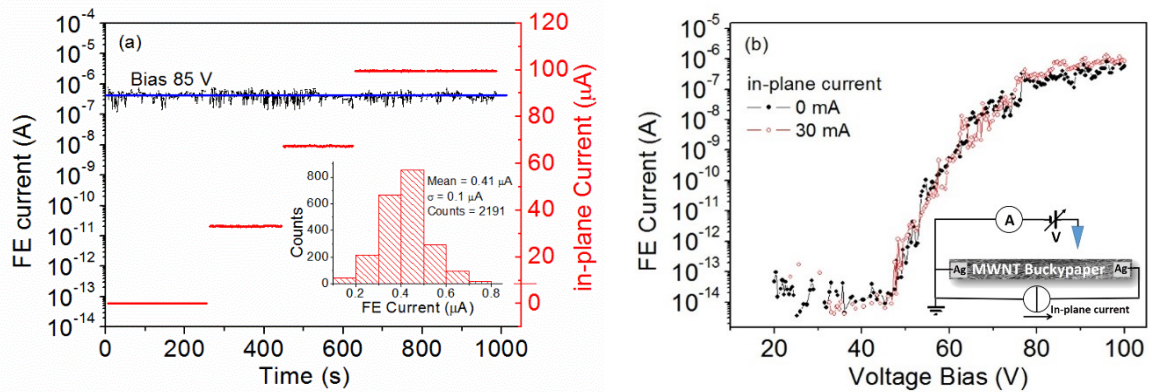

**Fig. 5 (a)** Time stability of the FE current for voltage bias of 85 V. The right axis (red) refers to the applied in-plane current. The measurement starts with zero in plane current and is raised up to 100 μA in three successive steps. Inset: Histogram gives statistic information about the FE current values recorded during the measurement. **(b)** Comparison of FE curves measured with (30 mA) and without in-plane current flowing in the buckypaper. Inset: schematic of the circuit to apply in-plane current during FE experiment.

We initially forced a current flowing in the buckypaper, along the alignment direction, in the range 0 – 100 μA by successive raising steps, while monitoring the FE current obtained at fixed bias of 85 V, as done for the usual stability experiment. In figure 5a we observe that the FE current remains insensible to the steps of the in-plane current. Moreover, we tried to apply higher currents (in the mA range) without significant modifications. We also compared the complete FE I-V characteristics measured with and without forced in-plane current in order to check if only limited bias regions could be affected by the presence of the in-plane current. However, experimental data reported in figure 5b show reproducible characteristics that confirm the absence of relevant effects of in-plane current on the spectra.

## Conclusions

We have fabricated CNT buckypaper using CVD grown aligned MWCNTs and we performed a deep experimental characterization of its transport and field emission properties. The temperature dependence of the conductance is described within the fluctuation induced tunneling model, but a further linear contribution is necessary in order to explain the behavior in the wide temperature range from 4.2 K to 430 K. We demonstrate that our buckypapers are extremely stable emitters with emitted current almost unaffected after half day operating time, reporting also turn-on fields as low as 140V/μm at distance below 1μm. We finally tested the possibility of FE tuning by forcing in-plane currents up to several mA in the buckypaper.

## Compliance with ethical standards

Conflict of interest. The authors declare that they have no conflict of interest.